\documentclass[11pt, a4paper]{article}
\pdfoutput=1
\usepackage{jinstpub}
\usepackage[utf8]{inputenc}
\usepackage{bookmark}
\usepackage[normalem]{ulem}
\usepackage{textcomp, gensymb}
\usepackage{graphicx}
\usepackage{siunitx}
\usepackage{natbib}
\usepackage{float}
\usepackage{appendix}
\usepackage{subfig}
\usepackage{wasysym}
\usepackage{comment}
\usepackage{parskip}
\usepackage{chemformula}
\usepackage{lineno}

\title{Scintillating properties of Cs${_2}$ZrCl${_6}$ crystals in the temperature range of 5--300~K}

\author[a]{F.~Cappella,}
\author[b]{P.C.F.~Di~Stefano,}
\author[b,c]{E.~Ellingwood,}
\author[b,c,d,**]{J.~Hucker,}
\author[b,c,e]{T.~Leroy,}
\author[b,c,f,g,*]{S.S.~Nagorny,}
\author[h]{V.V.~Nahorna,}
\author[f,g,b]{L.~Pagnanini,}
\author[b,c]{P.~Skensved,}
\author[b,c]{N.~Swidinsky,}
\author[h]{and P.~Wang}

\affiliation[a]{INFN, sezione di Roma, I-00185 Rome, Italy}

\affiliation[b]{Department of Physics, Engineering Physics and Astronomy, Queen's University, Kingston, ON, K7L 3N6, Canada}

\affiliation[c]{Arthur B. McDonald Canadian Astroparticle Physics Research Institute, Kingston, ON, K7L 3N6, Canada}

\affiliation[d]{Department of Physics, University of Toronto, Toronto, ON, M5S 1A7, Canada}

\affiliation[e]{Université de Lyon, Lyon, F-69361, France}

\affiliation[f]{Gran Sasso Science Institute, L'Aquila, I-67100, AQ, Italy}

\affiliation[g]{INFN -- Laboratori Nazionali del Gran Sasso, Assergi, I-67100, AQ, Italy}

\affiliation[h]{Department of Chemistry, Queen's University, Kingston, ON, K7L 3N6, Canada}

\affiliation[*]{Corresponding author: Serge Nagorny}
\emailAdd{serge.nagorny@gssi.it}

\affiliation[**]{Corresponding author: Jonathan Hucker}
\emailAdd{jon.hucker@utoronto.ca}


\abstract{A new comprehensive study on the Cs${_2}$ZrCl${_6}$ (CZC) crystal scintillating properties under different types of irradiation was performed over a wide temperature range from 5 to 300~K. The light yield (LY) at room temperature (RT), measured under irradiation by 662~keV $\gamma$ quanta of $^{137}$Cs, was evaluated to be 53,300 $\pm$ 4,700~photons/MeV corresponding to approximately 71\% of its estimated absolute value. The maximum light emission was observed in the temperature interval 135--165~K, where the LY reached 56,900~photons/MeV and 19,700~photons/MeV for $\gamma$ quanta and $\alpha$ particles, respectively. The quenching factor (QF) for $\alpha$ particles increases smoothly from QF~=~0.30 at RT to QF~=~0.36 at 135~K. The shape of scintillation pulses induced by $\alpha$ particles is characterized by three time-constants (0.3, 2.5 and 11.8~$\mu$s at RT), whereas the average pulse of $\gamma$ induced events is characterized by two time-constants (1.3 and 11.5~$\mu$s at RT). At the same time, scintillating properties and pulse-shape discrimination capability of the CZC exhibit an acute deterioration at temperatures below 135~K. The optimal operating conditions to maximize the scintillating performance of undoped CZC crystals are discussed.}

\begin{document}
\toccontinuoustrue
\maketitle
\section{Introduction}

At the present day, more than 500 inorganic scintillating compounds have been discovered and studied, and this list is continuously and extensively growing, see for instance~\cite{lawrence}. Due to their uniqueness and numerous possible usages, there is no single ideal scintillator and each scintillating material must be carefully characterized and selected to meet specific requirements of a certain application. Therefore, it is essential to understand their luminescence characteristics and material properties in a wide range of experimental conditions. For instance, in gamma ($\gamma$)-ray spectrometry there is a need for scintillators with a high light yield (near or above 100,000~photons/MeV), excellent energy resolution (better than 3\% at 662~keV $\gamma$ line of $^{137}$Cs), a high stopping power (Z$_{eff}$ larger than 70), and a fast decay time of scintillation (less than 1~$\mu$s). Physical and mechanical properties of a scintillating material and its overall cost, are also of importance, when considering the crystal for a large-scale application. For instance, the sodium iodide doped with thallium (NaI(Tl)) crystal scintillator was discovered in 1949~\cite{hofstadter} and has since been widely used in $\gamma$-ray spectrometry due to the combination of its low cost and acceptable scintillating performance. 
At the same time, scintillating materials used in fundamental studies typically possess some desirable unique features, which includes the possibility to introduce an element or isotope of interest into its chemical composition. Many such custom inorganic crystal scintillators with enriched isotopes have been developed over the last decades to search for rare nuclear processes such as rare beta ($\beta$) and alpha ($\alpha$) decays, neutrinoless double beta (0$\nu$2$\beta$) decay and search for axions. More details on various experimental techniques and crystal-based detectors used in this field can be found in~\cite{Agostini,zolotarova,danevich2,danevich1}. It follows that developing a crystal scintillator with tunable characteristics is a desirable feature.

Recently, a new family of bright inorganic crystal scintillators has been introduced~\cite{burger} and is the focus of extensive studies~\cite{nagorny1}. This is a series of metal hexachalcogenite compounds that can be described with a general formula A$_2$MX$_6$, where A is an element from alkali metals group (i.e. Li, Na, K, Rb, Cs) or Tl; X is a halogen element (F, Cl, Br, or I); and M is one of the tetravalent elements such as Zr, Sn, Te, Hf, Pt, Os, Re, Ru, etc. Accordingly, each element in the A$_2$MX$_6$ crystalline matrix can be substituted by its alternative. Overall, such variability in chemical composition makes this class of compound flexible with respect to the element or isotope of interest that can be embedded in light of its fundamental study, as well as to any possible adjustment of the crystal structure or scintillating characteristics required by a particular application. Moreover, this crystalline matrix could be doped with activators such as Ce$^{3^{+}}$ and Eu$^{3^{+}}$, or could be alloyed, which can further improve its scintillating performance~\cite{nagorny1,tang}. It should also be noted that the procedures in crystal production technology for compounds of this family are very similar. Consequently, progress in the production of high-quality crystals for scintillators of this crystal family, accumulation of this knowledge and experience, comprehensive investigation of their scintillating properties - all together prepare the basis for future successes in the study of various phenomena and processes in a wide variety of isotopes and elements.

In the framework of the recently executed CHARM (radiopure Crystal with HAfnium for Rare Mode of alpha decay searches) initiative at Queen's University (Kingston, Canada), Cs$_2$ZrCl$_6$ (CZC) crystals with dimensions of $\diameter$22 $\times$ 60~mm and mass of about 60~g were developed~\cite{belli1}. One of the main aim of this R\&D was to develop an appropriate detector material to study rare nuclear decays that may occur in natural Zr isotopes, such as the neutrinoless 2$\beta$ decay of $^{94,96}$Zr and strongly forbidden $\beta$ decay of $^{96}$Zr. The CZC compound is the first room-temperature ambient-tolerant crystal scintillator containing Zr in a relatively-high mass-fraction (16\%). It allows to implement the “source = detector” approach, when the decaying isotope is embedded in the detector material, that leads to a high detection efficiency of the processes of interest. Moreover, these detectors exhibit a high identification capability for $\gamma(\beta)$ and $\alpha$ particles, along with a reasonable energy resolution (Full Width at Half Maximum) FWHM = 5\% at 662~keV $\gamma$ line of a $^{137}$Cs source. Within long-term low-background measurements, novel experimental limits on the half-lives of various modes of the 2$\beta$ decays of $^{96}$Zr and $^{94}$Zr were set at the level of T$_{1/2}$ $\sim$ 10$^{17}$--10$^{20}$~years~\cite{belli1}. This was the first complex study of CZC scintillating crystals in terms of their chemical purity and internal radioactive contamination, detailed analyses of their internal background components and scintillating performance at RT that all together confirm the potential of these crystals as a detector for rare decays of Zr isotopes.

Soon after it was reported~\cite{swiderski} that the CZC crystal demonstrates a good linearity of the scintillation light response for internal electrons (electrons of the Compton scattering from external $\gamma$ quanta) in a wide energy range, from 30~keV up to 3~MeV. For external $\gamma$ quanta, the linearity is good down to about 100~keV. Below this energy, a non-proportionality of the light yield for external $\gamma$s is at the level of about 6\%.

For many scintillating crystals, the luminescence intensity increases as the crystal temperature decreases~\cite{mikhailik}. Commonly, the light output of undoped crystal scintillators increases evenly and continuously as the crystal temperature decreases or a step-like at a certain temperature, indicating that the material has passed the luminescence quenching temperature region which is characterized by the nature of luminescence centers and its internal defect structure. Recently, the temperature dependence of the scintillation light intensity and pulse time-constants were studied for the CZC crystal in the interval 4--300~K under irradiation by $\alpha$ particles and reported in~\cite{growth}. It was shown that CZC exhibits a unique behavior in the temperature region between 80~K and 125~K, where the intensity of the scintillation light decreases with the temperature decrease. This feature was attributed to the negative thermal quenching caused by self-trapped excitons (STE)~\cite{kang,koshimizu}. Since such behavior of the light yield has been reported only for $\alpha$ particles, further investigations of this phenomenon under $\gamma$s and X-rays irradiation are required. In the same study, the light yield for the CZC crystal was estimated as 33,900~photons/MeV at RT, when irradiated by 662~keV $\gamma$s originating from a $^{137}$Cs source~\cite{growth}. Because of such a non-trivial temperature dependence of the scintillation light yield, its maximum is observed not at the lowest temperature, but around 125~K. This allows one to establish an experiment based on CZC crystal scintillators at moderate temperatures while not compromising the experimental sensitivity and detector performance, such as energy resolution and pulse-shape discrimination capability.

In this study we present the results on the temperature dependence of the light yield of the CZC crystal scintillator measured over the temperature range 5--300~K under $\alpha$ particles and $\gamma$ quanta irradiation. The temperature dependence of scintillation pulse time-constants, their relative intensities, and quenching factor for $\alpha$ particles are also reported. An optimal operation temperature to achieve the maximum scintillation light yield and pulse-shape discrimination capability with CZC is discussed. All those aspects would be useful for planning future experiments with such scintillating detectors.

\section{Experimental}
\subsection{Crystal production}
The single crystal sample used in this study was obtained from Cs$_2$ZrCl$_6$ ingot ($\diameter$22 $\times$ 60~mm, 60~g) produced through a multi-stage crystallization process by the vertical Bridgman-Stockbarger technique adopted from~\cite{belli1}. In this method, CsCl (99.9\%) and ZrCl$_4$ (99.9\%) powders were used as starting materials. The latter was subjected to a two-stage sublimation process prior to the Cs$_2$ZrCl$_6$ compound synthesis. A pulling rate of 1.5~mm/hour and a temperature gradient of 25$^{\circ}$C/cm at the crystallization zone were maintained during the first crystal growth. Then, the obtained ingot was processed to remove all inclusions, as well as the first- and last-to-freeze sections. The second growth was performed with a 0.5~mm/hour pulling rate at the same temperature gradient. The final CZC single crystalline ingot was cut by a diamond wire saw in the way to obtain a non-oriented rectangular sample - 20 $\times$ 14 $\times$ 6~mm, which then was processed with a 1200 grit sandpaper and mineral oil as a lubricant. Before installation, the sample was wrapped with four layers of a Teflon tape (except of the side facing the PMT and area facing the $\alpha$ source) serving as a diffuse light reflector and as additional thermal insulation to prevent cracking of the crystal due to any possible sudden changes in temperature of the sample holder. Then it was secured in the sample holder using a few additional pieces of a Teflon tape.

\subsection{Table-top measurements at room temperature}

The crystal characterization was initially accomplished at RT with a $^{137}$Cs calibration source. The scintillator was coupled and secured directly to the entrance window of a photomultiplier tube (PMT, Electron Tubes 9125QSB) with optical grease. The coupled system was placed inside a dark-box to shield the setup from ambient light. The PMT was biased at a voltage of 900~V from a CAEN~N1470 power supply. The PMT output signal was transmitted to a NI PXIe~5162 digitizing unit for data processing, where the signal amplitude was measured in Analog-to-Digital Units (ADU). Each scintillating pulse was recorded over 400-$\mu$s-long time interval at a sampling rate of 200~MHz. The first 40~$\mu$s of the record was used to determine the baseline. To calculate the total charge, the pulse was integrated staring from 40~$\mu$s to 115~$\mu$s with baseline subtraction, and it was initially determined in units of ADU$\mu$s. A total of 10,000~events were acquired and analyzed offline.

\begin{figure}[H]
    \centering
    \includegraphics[width=0.7\textwidth]{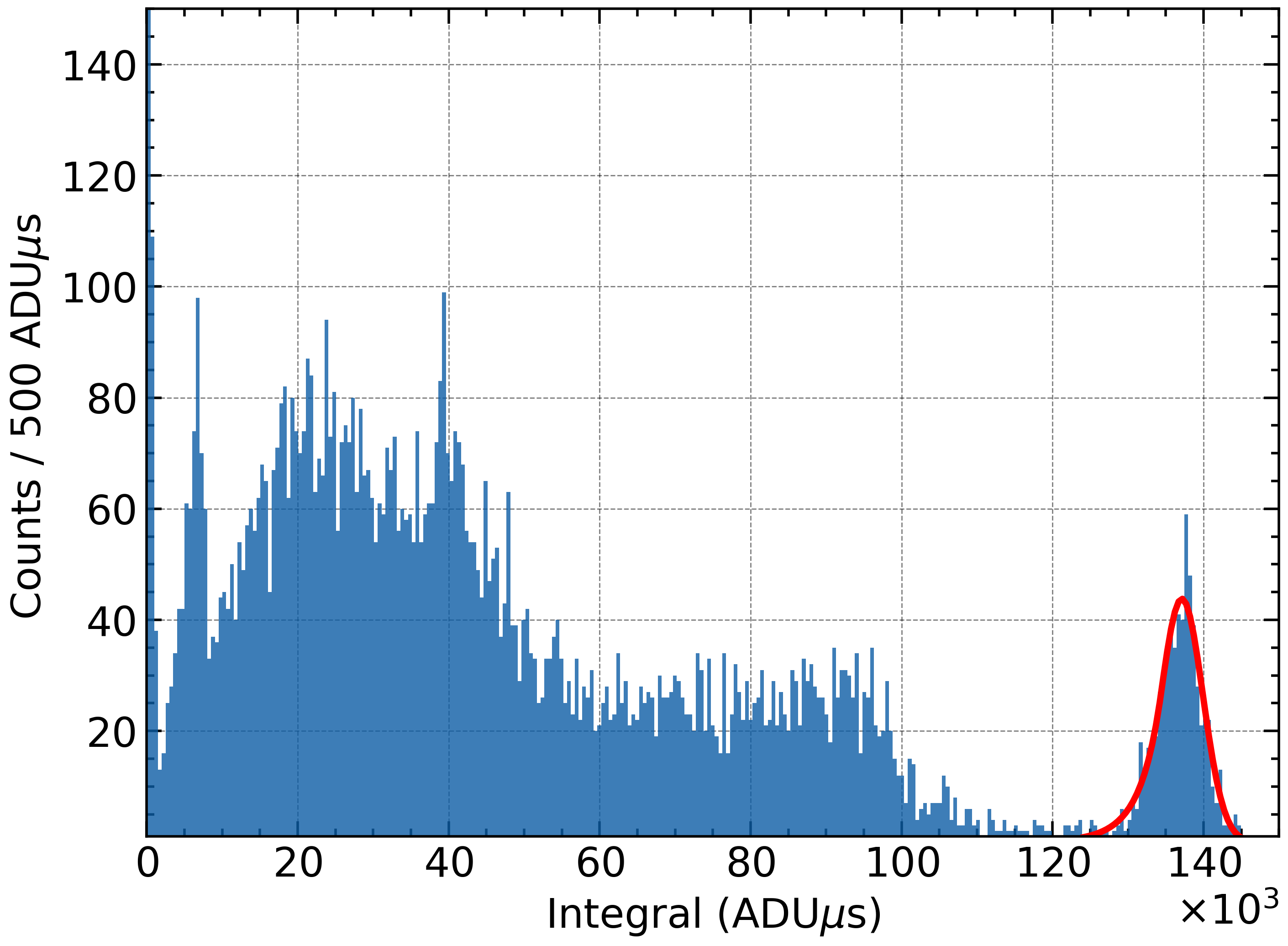}
    \caption{\label{Darkbox137Cs} Energy spectrum collected at RT with the CZC sample irradiated by 662~keV $\gamma$ quanta from a $^{137}$Cs source. Analyzing the spectrum from low to high energy, the narrow peak near zero corresponds to noise events with no corresponding scintillation light. The peak around 7,000~ADU$\mu$s is due to Ba unresolved X-rays (31.8 and 32.2~keV). The back-scattering peak and Compton edge at 40,000~ADU$\mu$s and 100,000~ADU$\mu$s, respectively, are present and well separated. The full absorption peak at 136,970~ADU$\mu$s is well defined with the energy resolution FWHM~=~4.7\%.}
\end{figure}

 The red curve seen in figure~\ref{Darkbox137Cs} corresponds to the fit of a piecewise continuous function, frequently referred to as a CrystalBall distribution, which consists of a Gaussian core and a power-law tail. The peak position (Integral$_{\rm{peak}}$) occurs at 136,970 $\pm$ 130~ADU$\mu$s units and standard deviation of 2,730 $\pm$ 100~ADU$\mu$s units. Statistical uncertainties corresponding to the best fit parameters are reported here. Systematic uncertainties are considered when reporting the experimentally evaluated light yield value. The binned negative log likelihood function provided in the Python package iminuit was minimized to obtain the best fit parameters of the peak.

The light yield (LY) of a scintillator could be determined as shown in eq.~(\ref{LY}) as the ratio between the number of photons produced (N$_{\rm{ph}}$) and the amount of energy deposited (E$_{\rm{\gamma}}$). The total charge generated at the anode of the PMT could be determined through the scintillation pulse integration over the data acquisition time-window.

\begin{align}
    \label{LY}
    LY({photons/MeV}) &= \frac{N_{ph}}{E_\gamma}\\
    \label{NPE}
    N_{ph} &= \frac{Integral_{peak}}{SPE}\frac{1}{QE\cdot LCE}
\end{align}

Where, the number of photons (N$_{\rm{ph}}$) was determined using  the peak position (Integral$_{\rm{peak}}$),  the charge generated at the anode of the PMT corresponding to a single photoelectron (SPE), the quantum efficiency of the PMT (QE), and the light collection efficiency of the setup (LCE).

It should be taken into account that the SPE value primarily depends on the PMT gain. Dedicated PMT calibration measurements were carried out using a 285~nm UV LED powered by a Kapustinsky pulsing circuit producing approximately 10~ns FWHM flashes at a repetition rate of 50~Hz~\cite{kapustinsky}. Charge collected by the PMT was integrated over a 50~ns integration interval with a sampling frequency of 1~GHz to obtain a distribution well characterized and fit by the single photoelectron distribution described by Tokar~et.~al~\cite{tokar}. This procedure returned an SPE of 42.3 $\pm$ 1.8~ADU$\rm\mu$s/photoelectron for the table-top experiment. In addition to the previous considerations discussed, the average charge measured at the anode of a PMT corresponding to the SPE depends on digitizer parameters, such as the vertical range and sampling frequency. Therefore, the quoted uncertainty includes not only a statistical component from the fit of the SPE distribution, but also a systematic part present due to adjustment for the differing sampling frequencies between the PMT calibration measurement and the table-top experiment. A similar PMT characterization and SPE determination was previously carried out and described in~\cite{corning1,hucker,ellingwood}. This calibration factor was used then to convert the ADUs into a number of photons.

Apart from the charge integral, there are many additional factors to consider when calculating the light yield of a scintillator. The QE of the PMT photocathode is a function of wavelength and hence the emission spectra of the scintillator must be accounted for. In this study, the QE of 12\% was evaluated by convolving the QE spectrum of the PMT photocathode~\cite{et} and the emission spectrum of the CZC crystal at RT~\cite{saeki}. A similar determination of the QE was carried out by Turtos et. al~\cite{lyso}. The uncertainty in the QE must account for a non-uniformity of the photocathode sensitive layer on the entrance window of the PMT. After communication with the PMT producer, for our particular case, a relative uncertainty of 5\% was assigned for the more uniform central part of the PMT window covering 80\% of its area, while the remaining 20\% of the window area, which is located near the window edge, have a higher relative uncertainty of 15\%. Thus, the total relative uncertainty of QE value at 7\% was found by taking the weighted average of both uncertainties.

Moreover, a fraction of light generated during a scintillation event is lost due to the geometry of the experiment~\cite{nagorny1}. This lost light can be quantified by the LCE of the experimental setup, and should be carefully determined as many experimental details can affect this parameter. To obtain a thorough understanding of the LCE, multiple Monte Carlo simulations were carried out with the Geant4 software package~\cite{agostinelli, allison1, allison2} (version 4.10.06.p01), using the G4OpticalPhysics constructor to activate all the optical physics processes. It includes, e.g., refraction and reflection at medium boundaries, bulk absorption, and Rayleigh scattering. Optical properties, such as refractive index and absorption length, of all mediums involved in the simulations are given as input parameters for these processes. In the simulations, the properties of the surfaces were also defined, since the behavior of optical photons depends on the nature and properties of the two materials that are joined at that particular boundary. Here, the processes at boundaries were determined using the UNIFIED model of optical surfaces~\cite{scintcount} implemented in Geant4. It applies to dielectric-dielectric interfaces and tries to provide a realistic simulation, which deals with all aspects of surface finish and reflector coating. All the surfaces (CZC, quartz windows, PMT's window and optical grease) are assumed to be polished (i.e., Snell's law is applied based on refractive index of the two media) except for the surface at the boundary between the CZC crystal and Teflon, where a fully Lambertian reflection is assumed as a reflection mechanism. The relative uncertainty of all simulated LCE values was estimated at 5\%.

It was determined that the LCE for $\gamma$ quanta in geometry where the CZC crystal was optically coupled directly to the window of the PMT was 81\%. Accounting for all above listed parameters, the total number of photons generated can be calculated from eqs.~(\ref{LY}) and~(\ref{NPE}). Hence, the light yield was experimentally evaluated to be 50,300 $\pm$ 4,800~photons/MeV under irradiation by 662~keV $\gamma$ quanta at RT.

\subsection{Measurements in optical cryostat}

Temperature controlled measurements were performed in a closed-cycle optical cryostat installed by ColdEdge Technologies. The cryocooler cold head was manufactured by Sumitomo Heavy Industries and capable of cooling the CZC sample to any temperature between 300~K to 3.5~K~\cite{verdier}. The compact geometry of the cryostat allows for optimization of the light collection efficiency and reduce stray light-leakage. Scintillation light traveling from a sample attached to the cold-finger passes through three Suprasil quartz windows, escaping the cryostat and then interacts with the photocathode of PMT. Moreover, the cryostat is housed in an acrylic glovebox to allow for careful control of the humidity during the sample installation. All external components were at RT that was monitored during the experiment.

\begin{figure}[htp]

\centering
\includegraphics[width=.45\textwidth, height = 9cm]{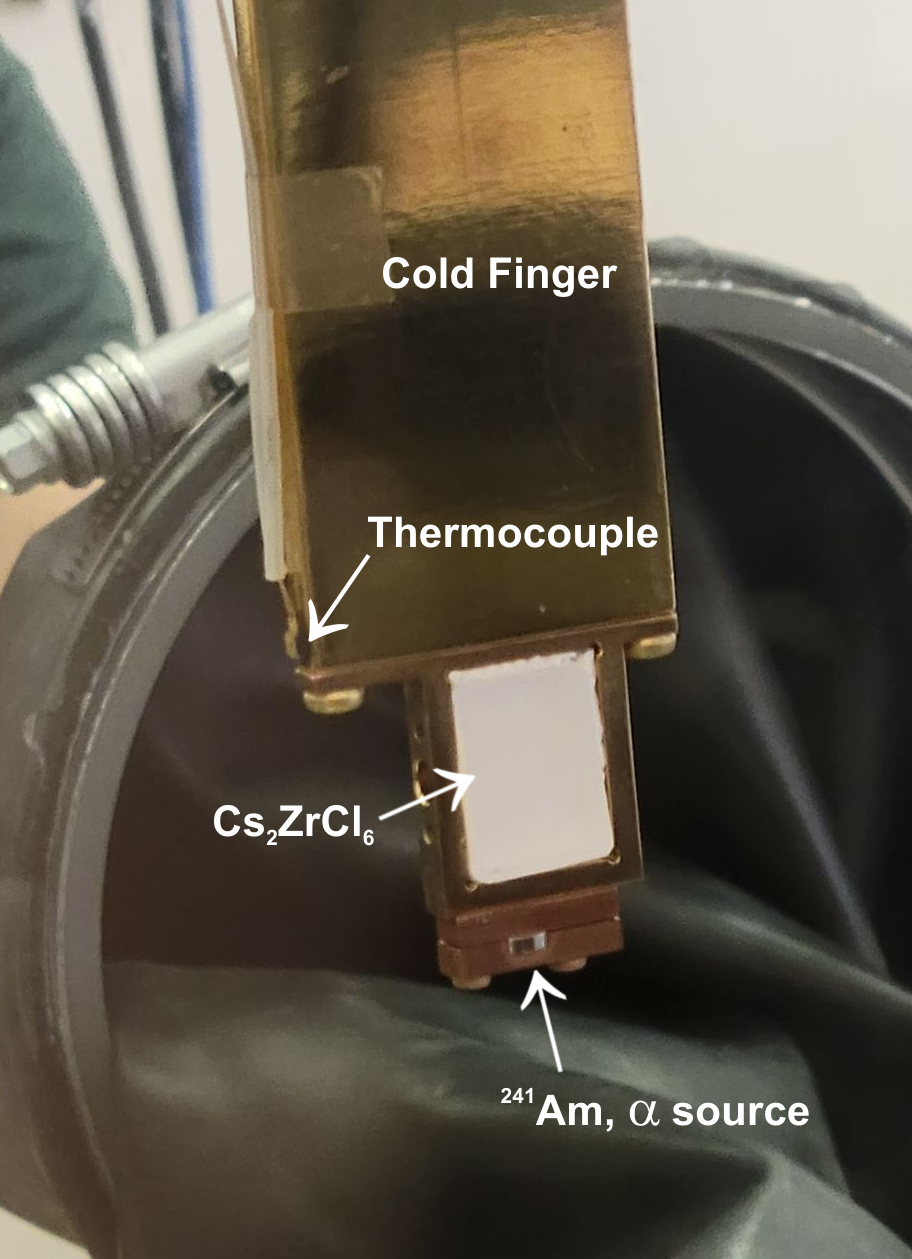}\hfill
\includegraphics[width=.45\textwidth, height = 9cm]{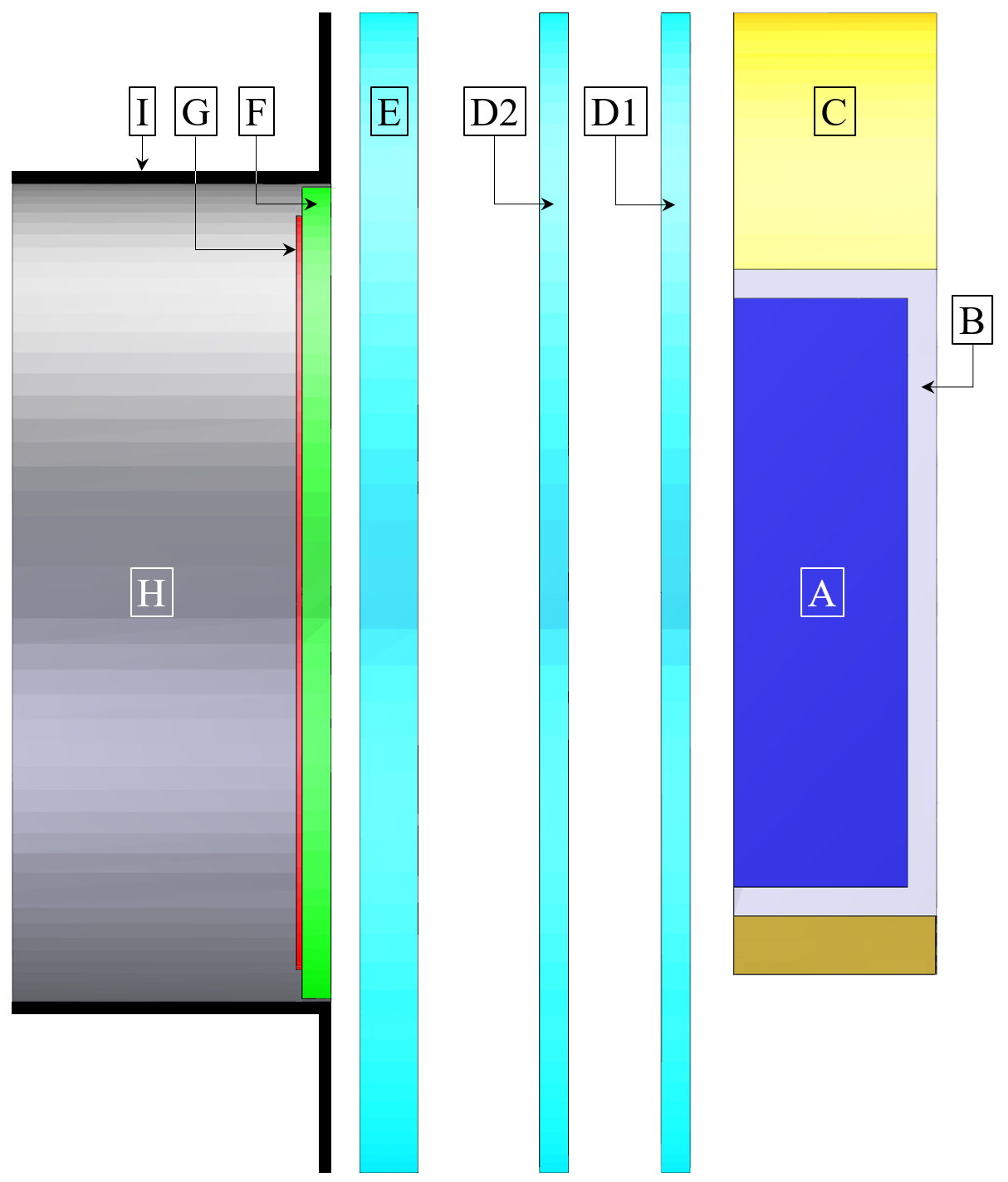}\hfill

\caption{(Left) The Cs${_2}$ZrCl${_6}$ crystal sample (20 $\times$ 14 $\times$ 6~mm) mounted in the crystal holder during installation at the cryostat. The thermocouple is mounted on a cold finger near the sample holder to ensure careful temperature monitoring. The $^{241}$Am $\alpha$ source, a small silver-colored tablet, can be seen at the bottom of the sample holder. (Right) The schematic view of the measurement setup with the Cs${_2}$ZrCl${_6}$ crystal installed in the optical cryostat used for the light collection efficiency (LCE) coefficient simulation in the Geant4 software package. Where: “A” is the crystal sample; “B” is the Teflon reflecting tape (1~mm thick); “C” is the gold-plated cold finger with an attached crystal holder; “D1” and “D2” are the fused quartz internal windows of the cryostat ($\diameter$40 $\times$ 1~mm); “E” is the fused quartz external window of the cryostat ($\diameter$40 $\times$ 2~mm); “F” is the fused quartz entrance window of the PMT ($\diameter$28 $\times$ 1~mm); “G” is the photocathode layer of the PMT; “H” is the PMT internal volume; “I” is the light-tight black-colored PMT holder that covers rest of the entrance window “E”. Area to the right of the window “E” represents internal volume of the optical cryostat, while area to the left of the window “E” stays at ambient pressure and temperature. Distances: “A-D1” = 1.5~mm, “D1-D2” = 3.2~mm, “D2-E” = 4.2~mm, “E-F” = 1.0~mm.}

\label{czcshroud}
\end{figure}

An external CryoCon-24 temperature controller system was used to set and monitor the temperature of the cold finger with the sample mounted in the crystal holder. A dedicated calibration run was performed to ensure the temperature of the cold finger was stabilized within 0.1~K of a desired temperature and that a steady cooling rate of 0.5~K/min was achieved.
The installation of the CZC sample was completed in a glove-box under a nitrogen atmosphere. A $^{241}$Am source (with a protective metal film that covers the radioactive material, reducing the energy of the alpha particles to 4.7~MeV) was secured underneath the crystal and can be seen in figure~\ref{czcshroud} (left). While the $^{241}$Am $\alpha$ source was in place for the entirety of the experiment, the $^{137}$Cs $\gamma$ source was placed just outside of the cryostat and can be removed at any time. The schematic view of the measurement setup with the CZC sample installed in the optical cryostat can be seen in figure~\ref{czcshroud} (right).

The data was collected in a single run, starting measurements at RT and ending at 5~K. This was done to prevent the possibility of charge carriers being released from shallow traps in the crystal lattice as a result of its heating, also known as a thermoluminescence phenomenon. A minimum delay of thirty minutes between reaching a desired temperature and data acquisition was maintained to ensure complete thermalization of the CZC crystal.

The scintillation light produced through particle interactions in the CZC sample was detected by the same PMT used in the table-top measurement. The high sampling frequency of 2.5~GHz, used in this set of measurements, ensured that individual photoelectron pulses were detected. At each temperature, 10,000 waveforms over 800-$\mu$s-long time window were recorded and analyzed offline. A typical energy spectrum collected with the CZC crystal mounted in the optical cryostat at RT under simultaneous irradiation by $^{241}$Am $\alpha$ source and $^{137}$Cs $\gamma$ source is presented in figure~\ref{RoomTEnergySpec}.

\section{Results}
\subsection{Light yield over 5--300~K temperature interval}

\begin{figure}[H]
   \centering
   \includegraphics[width=0.7\textwidth]{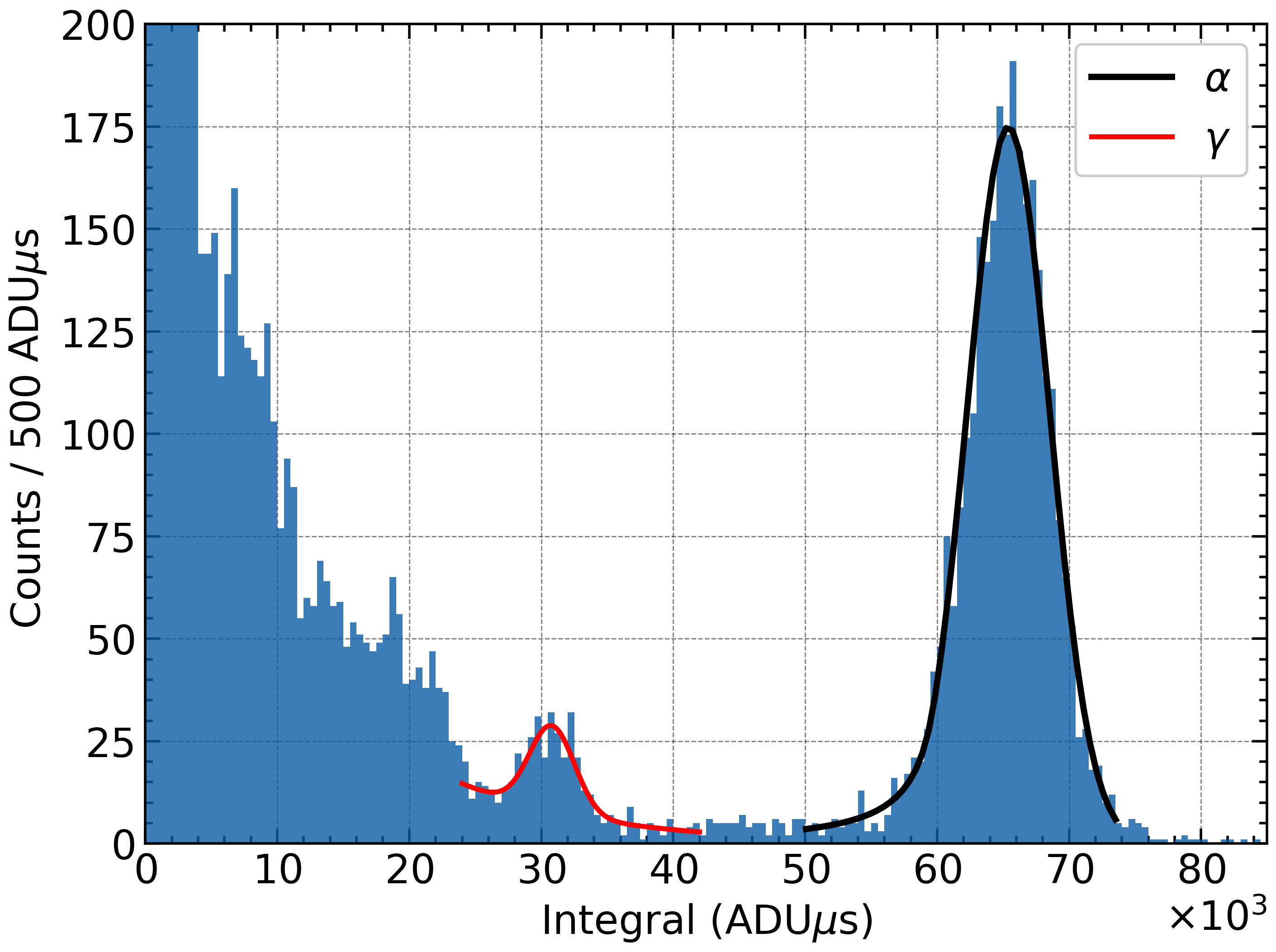}
    \caption{Energy spectrum collected at RT with the Cs${_2}$ZrCl${_6}$ crystal mounted in the optical cryostat. The red curve represents the best fit of the 662~keV full absorption peak from $^{137}$Cs with a sum of Gaussian and exponential functions. The black curve exhibits the best fit of the 4.7~MeV $\alpha$ peak from $^{241}$Am with the CrystalBall function. Integral is expressed here in ADU$\mu$s.}
    \label{RoomTEnergySpec}
\end{figure}

Energy spectra induced by $\alpha$ particles and $\gamma$ quanta in the CZC crystal installed in the optical cryostat were acquired over a series of temperatures from 295~K to 135~K. The energy spectrum collected at RT is shown in figure~\ref{RoomTEnergySpec}. Due to the more complex geometry of these measurements, less scintillation light can be collected to the PMT’s photocathode, resulting in a poorer energy resolution and a closer location of the 662~keV full absorption peak to the edge of the Compton scattering in case of $\gamma$ quanta detection. Hence, a sum of Gaussian and exponential functions was used to describe a 662~keV peak, residual background and some contribution of the edge of Compton scattering. While the CrystalBall function was used to determine the best fit parameters for the $\alpha$ peak, following the fitting procedure as described above in Section 2.2. The $\alpha$'s and $\gamma$'s energy resolutions corresponding to these fits were 11\% and 10\%, respectively. In addition, $\alpha$ energy spectra in the temperature range from 135~K to 5~K were acquired.

The mean of the fit provided a relative measurement of the light yield. The peak position of the $\gamma$'s was found to be 30,650 $\pm$ 250~ADU$\mu$s with a standard deviation of 1,800 $\pm$ 300~ADU$\mu$s while $\alpha$'s peak position was determined to be 65,400 $\pm$ 100~ADU$\mu$s with a standard deviation of 3,240 $\pm$ 40~ADU$\mu$s. The uncertainties provided for light yield measurements includes only statistical uncertainties associated with the best fit parameters.

From eqs.~(\ref{LY}) and~(\ref{NPE}) the LY could be determined, where SPE and QE were 30.2 $\pm$ 0.4~ADU$\rm\mu$s/pe and 0.12 $\pm$ 0.01, respectively. The LCE for $\alpha$ particles and $\gamma$ quanta were determined through independent Monte Carlo simulations of the CZC sample in the cryostat geometry to be the same 24\% for both types of irradiation. The relative uncertainty of the LCE values was estimated at 5\%. For example, substituting the parameters from the 295~K measurement into eqs.~(\ref{LY}) and~(\ref{NPE}), the LY for $\alpha$ particles and $\gamma$ quanta were determined to be 16,000 $\pm$ 1,400~photons/MeV and 53,300 $\pm$ 4,700~photons/MeV, respectively. The Quenching Factor (QF), defined as the LY ratio of $\alpha$ particles to $\gamma$ quanta for the same deposited energy, was determined to be 0.30 $\pm$ 0.04.

\begin{figure}[H]
   \centering
    \includegraphics[width=0.7\textwidth]{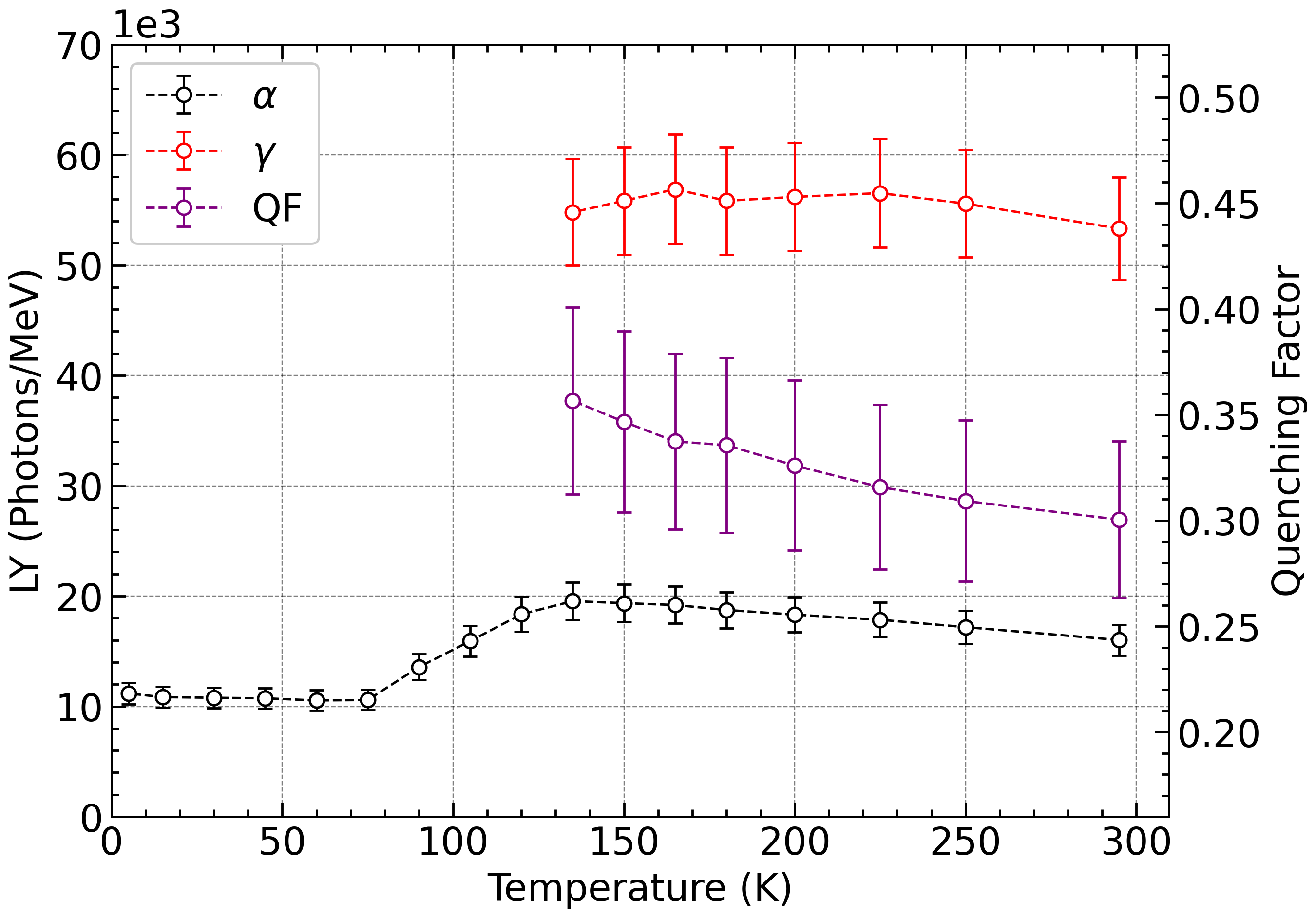}
    \caption{
    Temperature dependence of the Light Yield (LY) and Quenching Factor (QF, purple curve) of the Cs${_2}$ZrCl${_6}$ crystal under excitation by $\gamma$ quanta (red curve) and $\alpha$ particles (black curve). The LY is expressed in photons/MeV.}
    
    \label{LYandQFvsT}
\end{figure}

The same fitting procedure as performed at 295~K was repeated for all temperature points and its result is illustrated in figure~\ref{LYandQFvsT}. The QF measured for 4.7~MeV $\alpha$ particle irradiation, increased from 0.30 at RT up to 0.36 at 135~K. Two distinct regions are identified. Between RT and 180~K, the QF slowly and steadily increases, corresponding to a steady rise in LY$_{\alpha}$ and near constant LY$_{\gamma}$. Between 180~K and 135~K, the QF begins to increase more rapidly, corresponding to the reduction in LY$_{\gamma}$ in this temperature interval. The QF measured with the CZC crystal at RT is in a perfect agreement with QF values determined for CZC crystals in~\cite{belli1,belli2} and is slightly lesser compared to the Cs${_2}$HfCl${_6}$ (CHC) crystal with a QF of 0.41 for 4.7~MeV $\alpha$ particles~\cite{belli3}.

\subsection{Pulse-shape analysis}

It is known for many scintillators that the shape of the scintillation pulse varies for different types of irradiation used for its excitation, reflecting different physics processes in the crystal lattice responsible for the scintillation pulse formation.

Initially, events with a noisy baseline, pileups, and badly determined staring point were discarded from the collected dataset. Then, the different peak position on the detector energy scale for events induced by $\alpha$ and $\beta/\gamma$ irradiation was used to preliminary distinguish and select events of each type. See, for instance, the corresponding peak positions on the energy spectrum shown in figure~\ref{RoomTEnergySpec}. Events within two sigma around mean value of the corresponding peaks were selected for the further analysis. Only events with energy corresponding to a 662~keV full absorption peak were further considered as $\gamma$ events in the pulse-shape analysis. Next, the mean time and prompt-to-total collected charge ratio were calculated for each event, and were used as additional discrimination and selection parameters to improve quality of the selected “pure” event of each type. These methods are well described in~\cite{cardenas}, where authors adopted those techniques to discriminate signals induced by $\alpha$ particles from those of $\gamma$ quanta in measurements with a CHC crystal.

\begin{figure}
    \begin{minipage}[c]{\textwidth}
        \begin{minipage}[t]{0.5\textwidth}
            \centering
            {\includegraphics[height=0.7\textwidth]{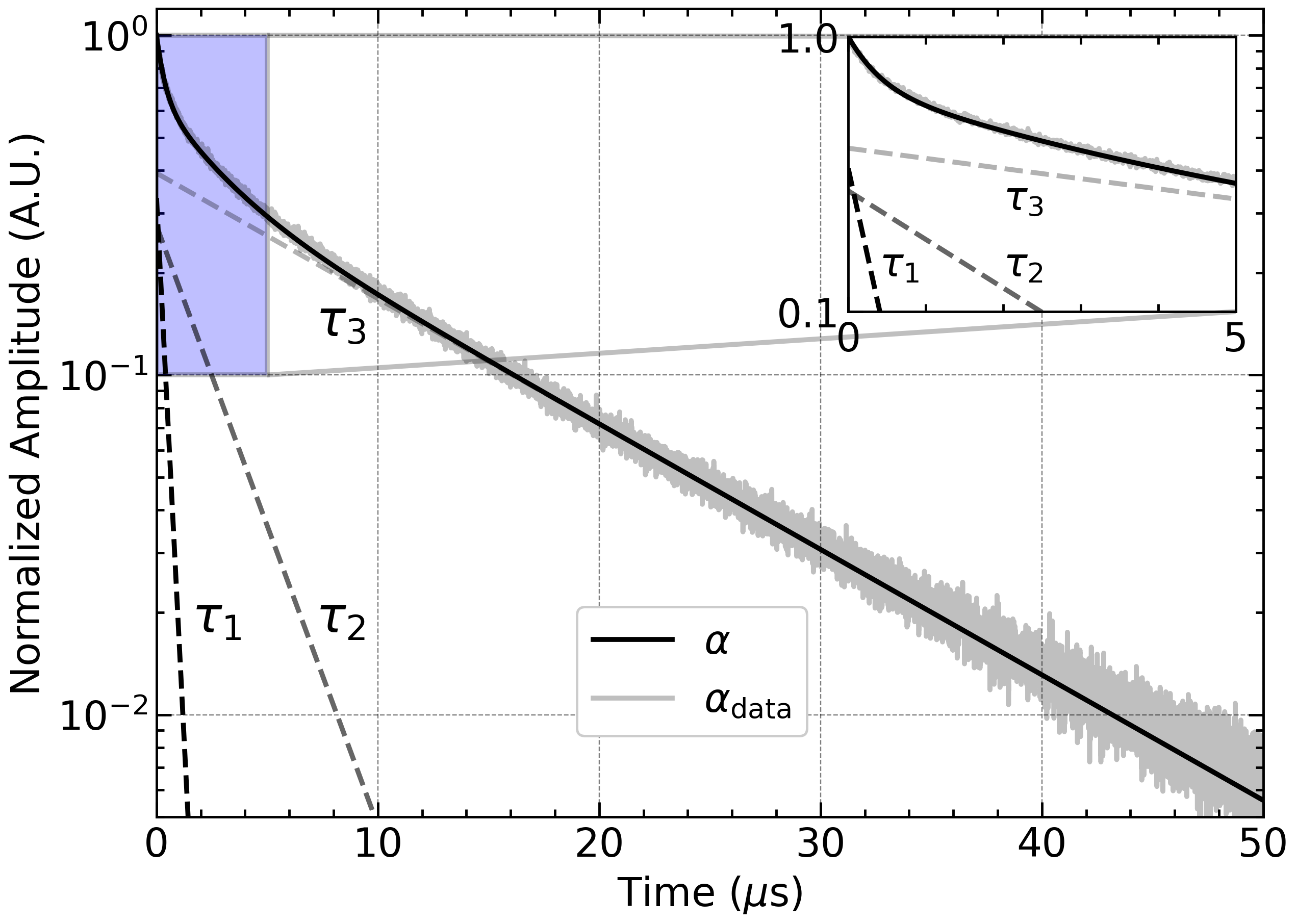}}
        \end{minipage}
        \begin{minipage}[t]{0.5\textwidth}
            \centering
            {\includegraphics[height=0.7\textwidth]{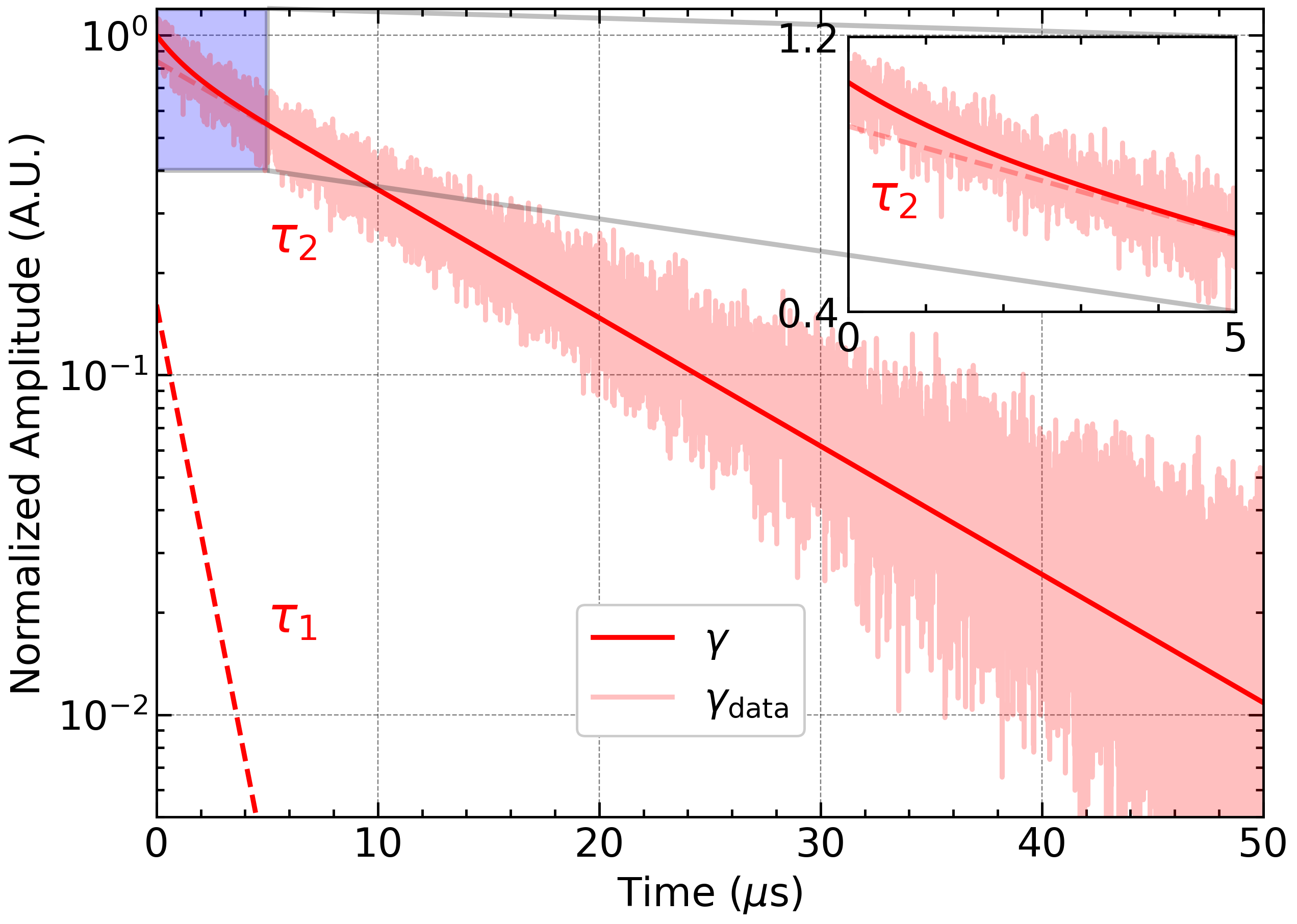}}
        \end{minipage}
    \end{minipage}
    \caption{\label{AvsGfitnormlog}Best fit curves of average $\alpha$ and $\gamma$ events recorded at RT are overlaid on the actual average pulses. The average pulses of each type were obtained through the addition of 1324 and 415 individual pulses induced by $\alpha$ (black curve) and $\gamma$ (red curve) events, respectively. These events were selected near the mean value of the corresponding peaks, and normalizing them to the same area.
    }
\end{figure}

Once “pure” events of each irradiation type were identified, the corresponding average waveforms were calculated. Then, the average $\alpha$ waveform was fitted with a triple exponential function described by eq.~\ref{AlphaEq}, where $I_1+I_2+I_3 =1$, as previously shown in~\cite{growth}. To fit the average $\gamma$'s waveform was used a double exponential function provided by eq.~\ref{GammaEq}, where $I_1 + I_2 = 1$. In both eq.~\ref{AlphaEq} and eq.~\ref{GammaEq}, $t$ represents the time duration of the average pulse, $\tau_i$ is the decay time-constant, and $I_i$ is the normalized relative intensity. The best fit of both types average waveforms normalized to their area, see figure~\ref{AvsGfitnormlog}, indicates that the average $\gamma$'s waveform lasts longer than the average $\alpha$'s waveform.

\begin{equation}
    f(t;\tau_1,\tau_2,\tau_3) = I_1\exp{(-t/\tau_1)}+I_2\exp{(-t/\tau_2)}+I_3\exp{(-t/\tau_3)}
    \label{AlphaEq}
\end{equation}
\begin{equation}
    f(t;\tau_1,\tau_2) = I_1\exp{(-t/\tau_1)}+I_2\exp{(-t/\tau_2)}
    \label{GammaEq}
\end{equation}

\begin{figure}
    \begin{minipage}[c]{\textwidth}
        \begin{minipage}[t]{0.5\textwidth}
            \centering
            {\includegraphics[height=\textwidth]{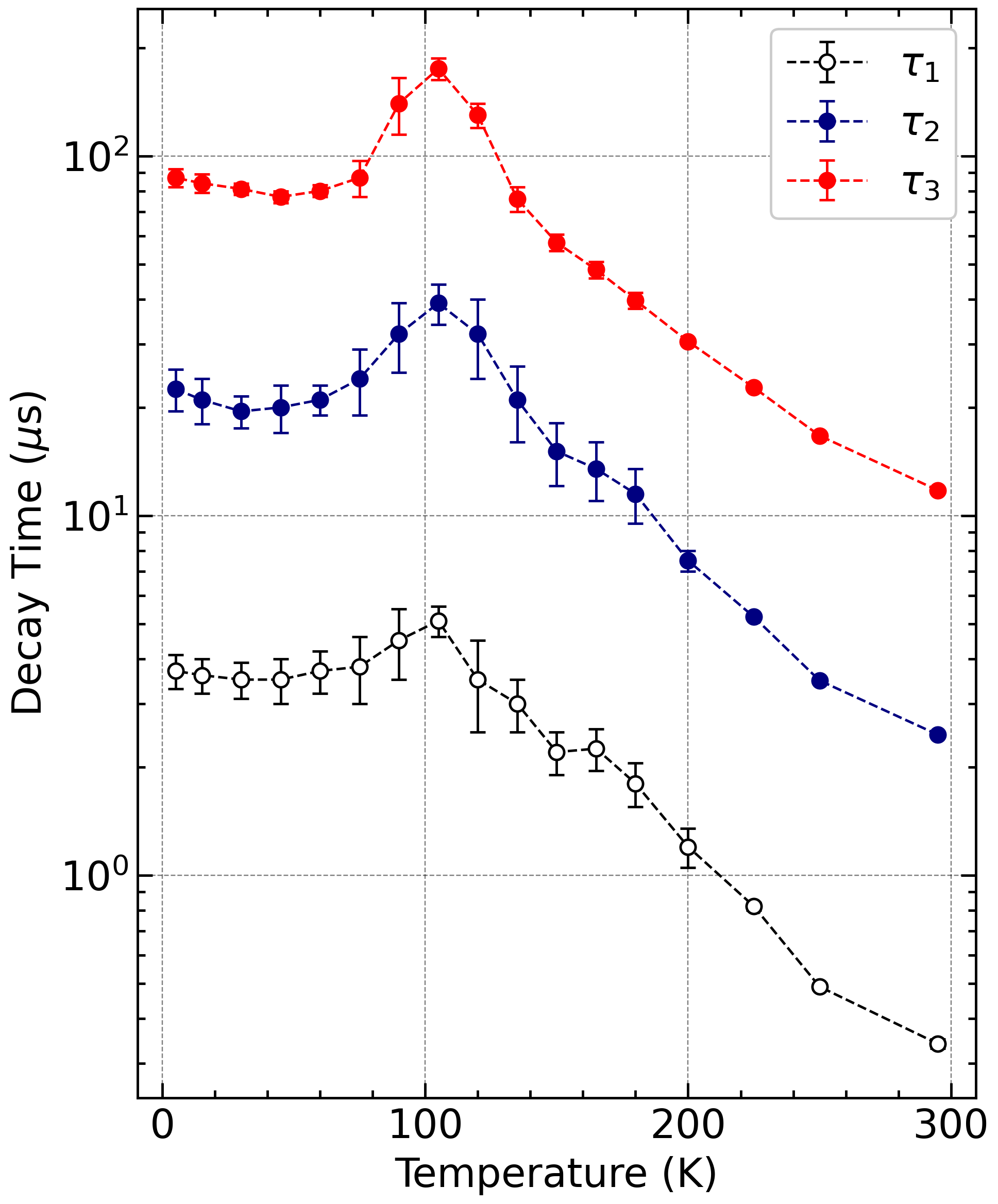}}
        \end{minipage}
        \hfill
        \begin{minipage}[t]{0.5\textwidth}
            \centering
            {\includegraphics[height=\textwidth]{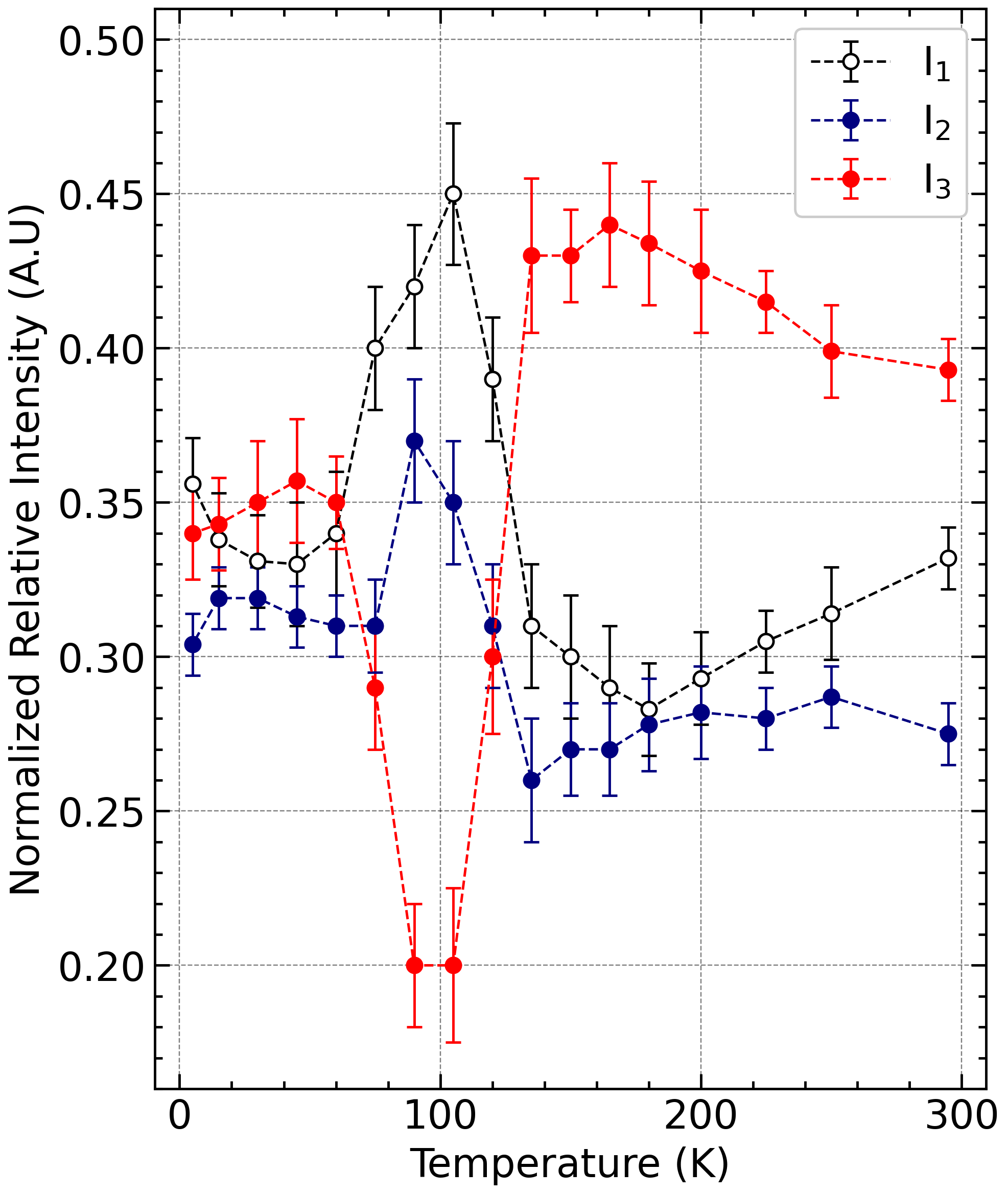}}
        \end{minipage}
    \end{minipage}
    \caption{\label{AverageAlphaTauAmp}Temperature dependence of decay time-constants (left) and their corresponding normalized relative intensities (right) for scintillating pulses induced by 4.7~MeV $\alpha$ particles in the Cs${_2}$ZrCl${_6}$ crystal, and obtained by fitting the shape of scintillating pulse averaged over 1000 individual “pure” $\alpha$ events with a sum of three exponential functions.}
\end{figure}

\begin{figure}[H]
    \begin{minipage}[c]{\textwidth}
        \begin{minipage}[t]{0.5\textwidth}
            \centering
            {\includegraphics[height=\textwidth]{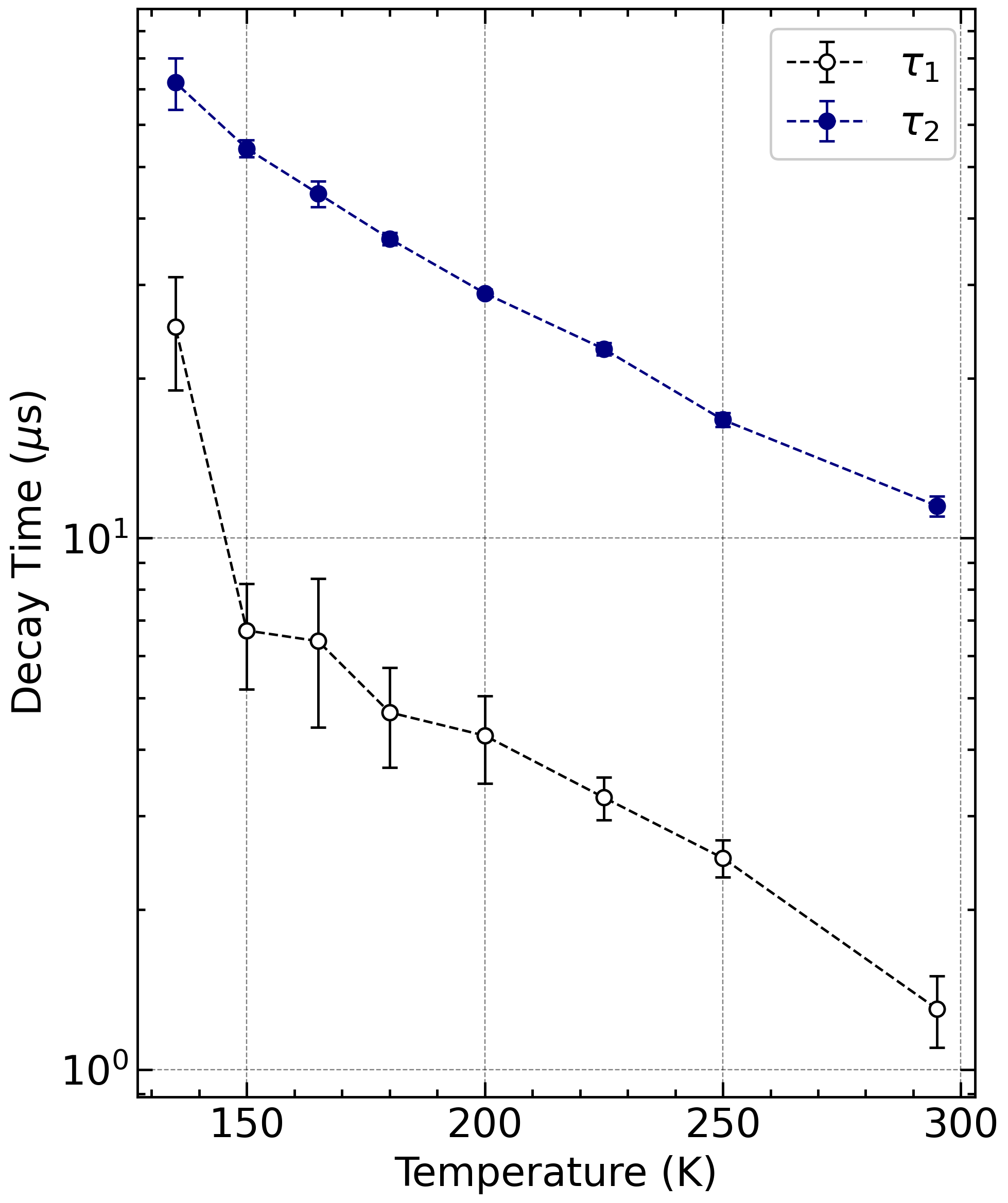}}
        \end{minipage}
        \hfill
        \begin{minipage}[t]{0.5\textwidth}
            \centering
            {\includegraphics[height=\textwidth]{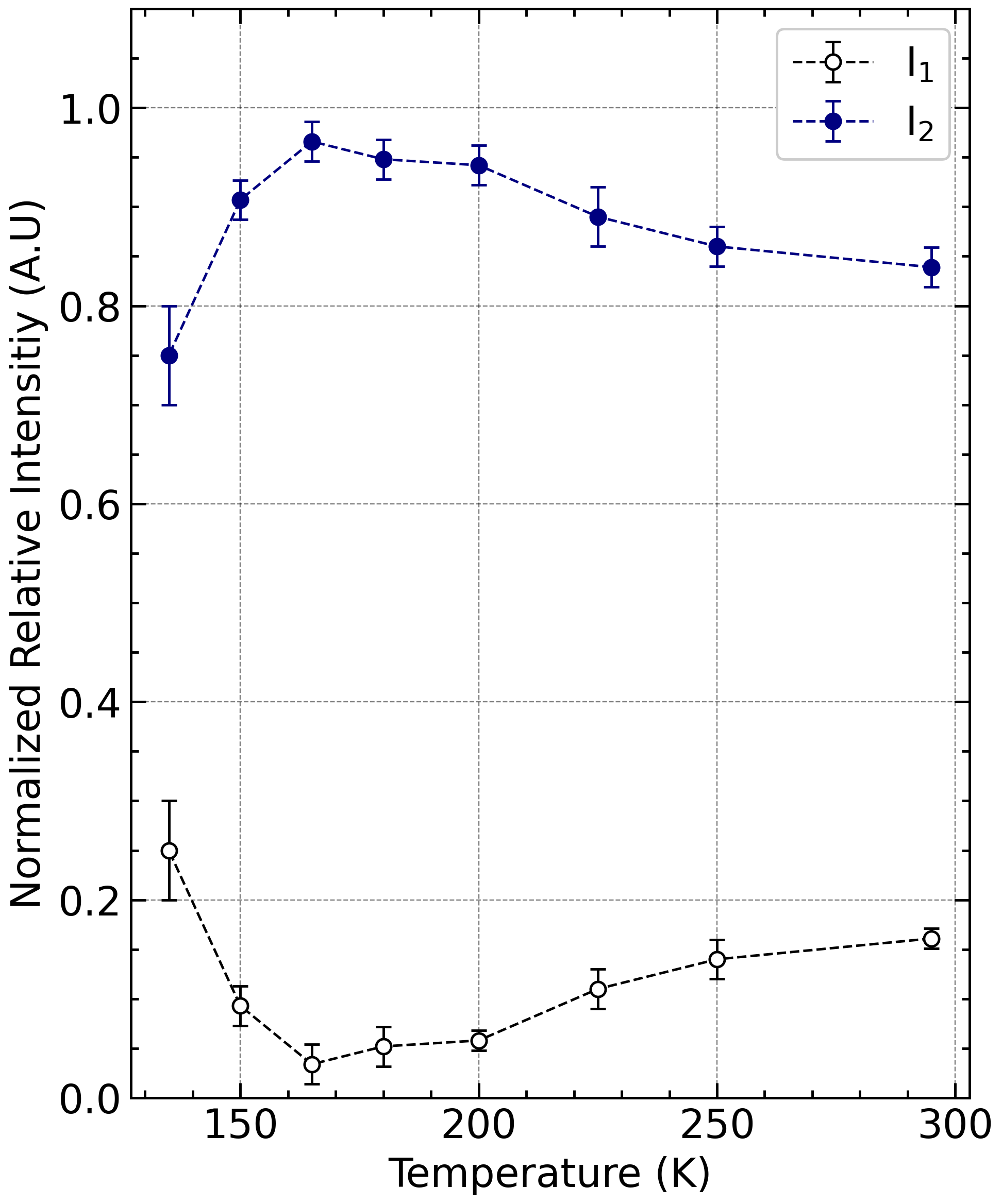}}
        \end{minipage}
    \end{minipage}
    \caption{\label{AverageGammaTauAmp}Temperature dependence of decay time-constants (left) and their corresponding normalized relative intensities (right) for scintillating pulses induced by 662~keV $\gamma$ quanta in the Cs${_2}$ZrCl${_6}$ crystal, and obtained by fitting the shape of scintillating pulse averaged over 500 individual “pure” $\gamma$ events by a sum of two exponential functions.}
\end{figure}

This pulse-shape analysis was performed at all temperatures studied for $\alpha$ events, but only down to 135~K for $\gamma$s, see figures~\ref{AverageAlphaTauAmp} and ~\ref{AverageGammaTauAmp}, respectively. Limitations of the current experimental setup (above ground location, permanently installed $\alpha$ source) prevented the study with $\gamma$ events down to lower temperatures. This was mainly due to: a) an overlap of scintillating pulses induced by these two types of calibration sources; b) an overlap with pulses induced by environmental $\gamma$ radioactivity and cosmic muons; c) a larger impact of a low-frequency baseline fluctuations on determining the starting point of scintillation pulses; and d) a relatively narrow time-window available (800~$\mu$s) for pulses recording, that was not enough to record the full length of the scintillation pulse at temperatures below 130~K. The long-lasting average decay time of scintillation pulses makes it necessary to perform further studies in a shielded optical cryostat located in an underground laboratory with a reduced environment radioactivity, since that would decrease the background counting rate in the crystal.

It should be emphasized that measured temperature dependence of the decay time-constants and their relative intensities for scintillating pulses induced by $\alpha$ particles in the Cs${_2}$ZrCl${_6}$ crystal, is confirming the behavior of those parameters observed earlier with such type of crystal in~\cite{growth}, see figure~\ref{AverageAlphaTauAmp}. From 5~K to about 70~K, the decay time-constants do not change significantly. In the temperature interval from 70~K to 140~K, all three decay time-constants experience an abrupt increase, and then from 150~K to RT an almost smooth reduction of all three decay time-components are observed. More complex dynamics were detected for normalized relative intensities of these decay time-components. In the interval from 5~K to about 70~K, the normalized relative intensities do not change significantly, following the trend observed for their decay time-constants. However, an abrupt redistribution occurs between 70~K and 140~K, leading to a minor contribution of the slowest time-constant to the pulse profile (only about 20\%) and increasing the contribution of the fast and medium decay time-constants. While above 140~K, the slowest time-constant is responsible for a 40--45\% of the total intensity of the scintillation pulse. As was demonstrated in~\cite{growth}, this significant change of decay time-constants and redistribution of their normalized relative intensities could be attributed to the thermal activation of certain emission centers, followed by the release of the trapped charges. This not only increases the scintillation intensity, but also slows down the scintillating emission. Above 140~K, the emission rate is accelerating again due to the thermal quenching.

The temperature dependence of the decay time-constants and their normalized relative intensities for scintillating pulses induced by $\gamma$ quanta follows the same trend as for pulses induced by $\alpha$ particles, being measured from RT down to 135~K, see figure~\ref{AverageGammaTauAmp}. The short component is responsible only for 3.5--14\% of the total intensity of the scintillation pulse in the 150--250~K temperature interval, reaching minimum at 165~K. However, below 150~K, the relative contribution of the short and slow decay time-components to the intensity of scintillation emission begins to be re-distributed.

As was already mentioned, the scintillation pulses decay curves were fitted using three (or two for $\gamma$ events) exponential functions that ensure the best quality of the fit. The reduced $\chi^2$-statistics ($\chi^2/\rm{d.o.f.}$, where d.o.f. is the number of degrees of freedom) was determined for $\alpha$ and $\gamma$ events at RT to be 261956/212461 and 69811/59587, respectively, which yield similar values of approximately 1.2. It should be noted though, that in the case of a complex decay process, such fits are merely a way to quantify the measured decay curves, and the fit parameters are not directly related to the specific emission processes occurring in the scintillation material. A precise determination of the involved luminescence centers, the nature and dynamics of emission processes were not the scope of the current study, and requires a dedicated complex investigation. Below, we listed decay time-constants and their corresponding normalized relative intensities for scintillating pulses induced by $\alpha$ particles and $\gamma$ quanta at certain temperatures, see table~\ref{valuesalpha} and table~\ref{valuesgamma}, respectively. These numerical data, in addition to the above listed plots, will help reader to better understand the temperature trend of scintillation emission for different types of irradiation.

\begin{table}[ht]
\begin{center}
\caption{\label{valuesalpha}Best fit decay time-constants and their corresponding normalized relative intensities for $\alpha$ induced scintillation events at selected temperatures.}
\begin{tabular}{|c | c | c | c | c | c | c |} 
 \hline
  T,~K & $\tau_1$,~$\mu$s & $I_1$ &$\tau_2$,~$\mu$s & $I_2$ & $\tau_3$,~$\mu$s & $I_3$  \\ [0.5ex] 
 \hline\hline
 295 & 0.3(1) & 0.33(1) & 2.5(2) & 0.28(1) & 11.8(2) & 0.39(1) \\
 \hline
 250 & 0.5(1) & 0.31(2) & 3.8(2) & 0.29(2) & 16.7(2) & 0.40(2) \\ 
 \hline
 200 & 1.3(2) & 0.33(1) & 8.5(5) & 0.30(1) & 30.5(5) & 0.37(2) \\
 \hline
 165 & 2.3(3) & 0.29(2) & 13.5(25) & 0.27(2) & 48.3(25) & 0.44(2)\\
 \hline
 135 & 3.0(5) & 0.31(2) & 21(5) & 0.26(2) & 76(6) & 0.43(2) \\
 \hline
 105 & 5.1(5) & 0.45(2) & 39(5) & 0.35(2) & 175(12) & 0.20(3) \\
 \hline
 15 & 3.6(4) & 0.34(2) & 21(3) & 0.32(1) & 84(5) & 0.34(2) \\ [0.5ex] 
\hline
\end{tabular}
\end{center}
\end{table}

\begin{table}[ht]
\begin{center}
\caption{\label{valuesgamma}Best fit decay time-constants and their corresponding normalized relative intensities for $\gamma$ quanta induced scintillation events at selected temperatures.}
\begin{tabular}{| c | c | c | c | c |} 
 \hline
 T,~K & $\tau_1$,~$\mu$s & $I_1$ & $\tau_2$,~$\mu$s & $I_2$   \\ [0.5ex] 
 \hline\hline
 295 & 1.3(2) & 0.16(2) & 11.5(5) & 0.84(2) \\
 \hline
 250 & 2.5(2) & 0.14(2) & 16.7(5) & 0.86(2) \\ 
 \hline
 200 & 4.3(8) & 0.06(1) & 28.9(5) & 0.94(2)\\
 \hline 
 165 & 6.4(20) & 0.03(2) & 44.5(25) & 0.97(2) \\
 \hline
 135 & 25(6) & 0.25(5) & 72(8) & 0.75(5)\\ [0.5ex] 
 \hline
\end{tabular}
\end{center}
\end{table}

\subsection{Temperature dependent pulse-shape discrimination}

\begin{figure}[H]
    \centering
    \includegraphics[width=0.7\textwidth]{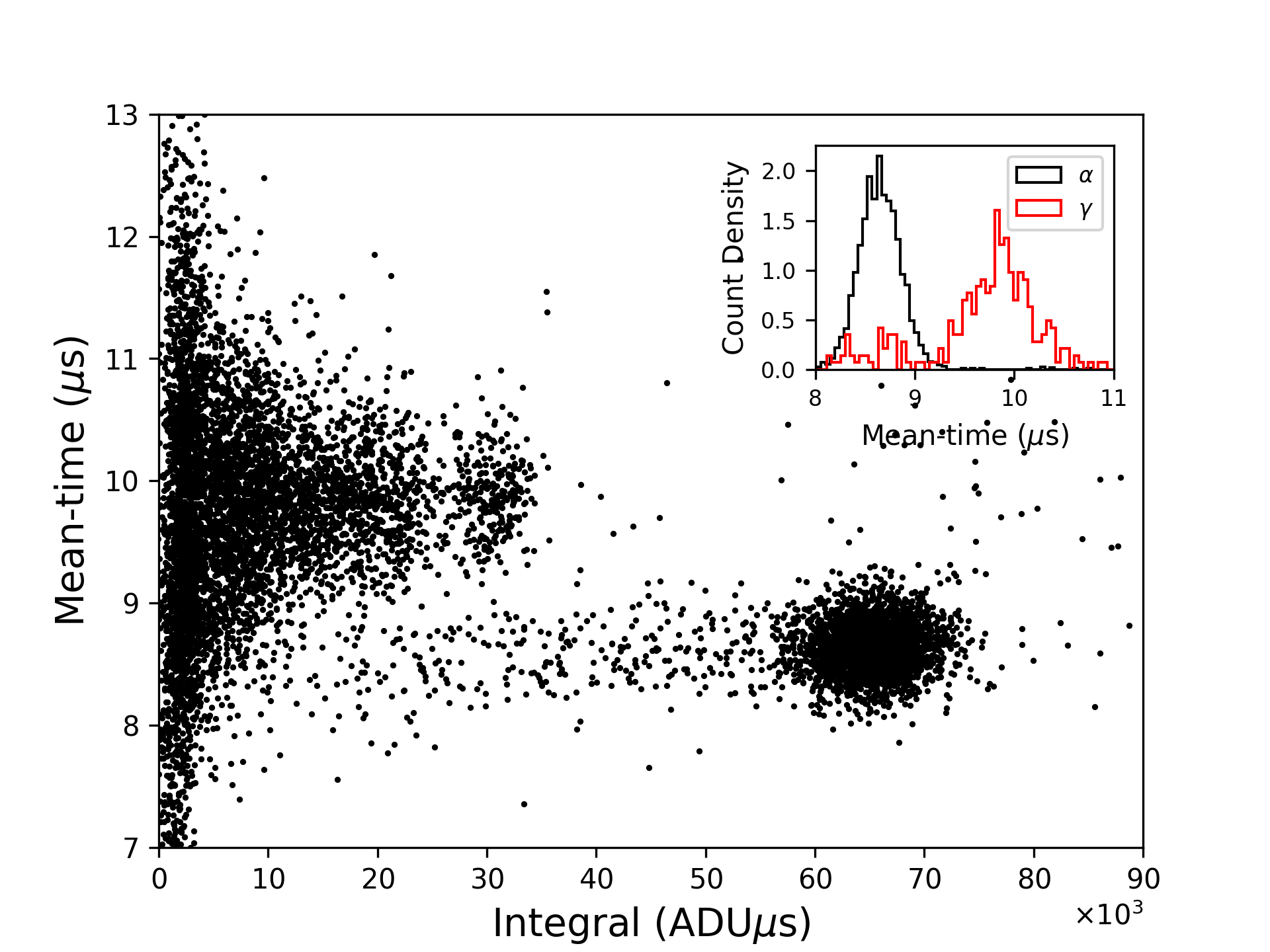}
    \caption{\label{PSD} Mean-time vs energy scatter plot recorded under simultaneous irradiation by $\alpha$ particles and $\gamma$ quanta with the Cs${_2}$ZrCl${_6}$ crystal installed in the optical cryostat at RT. There are two clearly visible and separated event distribution related to $\alpha$ particles and $\gamma$ quanta. Energy is reported here in ADU$\mu$s. Inset displays the mean-time distribution corresponding to $\alpha$ (black curve) and $\gamma$ (red curve) events over the energy interval 20,000--80,000~ADU$\mu$s, no additional selection on the individual events were performed.}
\end{figure}

Variation in scintillation pulses time-profile induced by particles of different types can be used to distinguish between them and to separate useful events from spurious and/or background events. The pulse-shape discrimination (PSD) technique could be realized through various methods: Optimum Filter~\cite{gatti}, mean-time~\cite{bardelli,vuong,belli1}, charge integration method~\cite{vuong}, etc. In the recent article by Cardenas~et.~al.~\cite{cardenas}, it was shown that all of those could be effectively applied to crystals from the CHC-like crystal family. In this work, the technique based on the scintillation pulse mean-time (see e.g. \cite{cardenas}), was adopted to discriminate $\beta(\gamma)$ induced events (662~keV $\gamma$ quanta of $^{137}$Cs) from events caused by 4.7~MeV $\alpha$ particles of $^{241}$Am calibration source. In particular, the time-profile of each scintillating event is exploited to calculate its mean-time according to the formulae: 

\begin{eqnarray}
\langle t \rangle = \left.\sum f(t_k) t_k\right/\sum f(t_k)
\end{eqnarray}

where the sum is taken over the time channels, $k$, starting from the origin of the pulse up to 70~$\mu$s at RT. Moreover, $f(t)$ is the digitized amplitude (at the time $t$) of a given pulse. To account for the possible elongation of scintillation pulses at lower temperatures, the mean-time calculation window was chosen at each temperature to be at least three times the length of the longest decay time-constant. For example, the mean-time calculation window was 70~$\mu$s at 295~K and was increased to a maximum of 300~$\mu$s at 135~K. The scatter plot of the mean-time versus the energy for the collected data with the Cs${_2}$ZrCl${_6}$ crystal at RT is shown in figure~\ref{PSD}. As one can see, the crystal exhibits a good pulse-shape discrimination capability for $\alpha$ particles (low mean-time population of events) and $\gamma$ quanta (high mean-time population of events) over the entire energy interval.

\begin{figure}[H]
    \centering
    \includegraphics[width=0.7\textwidth]{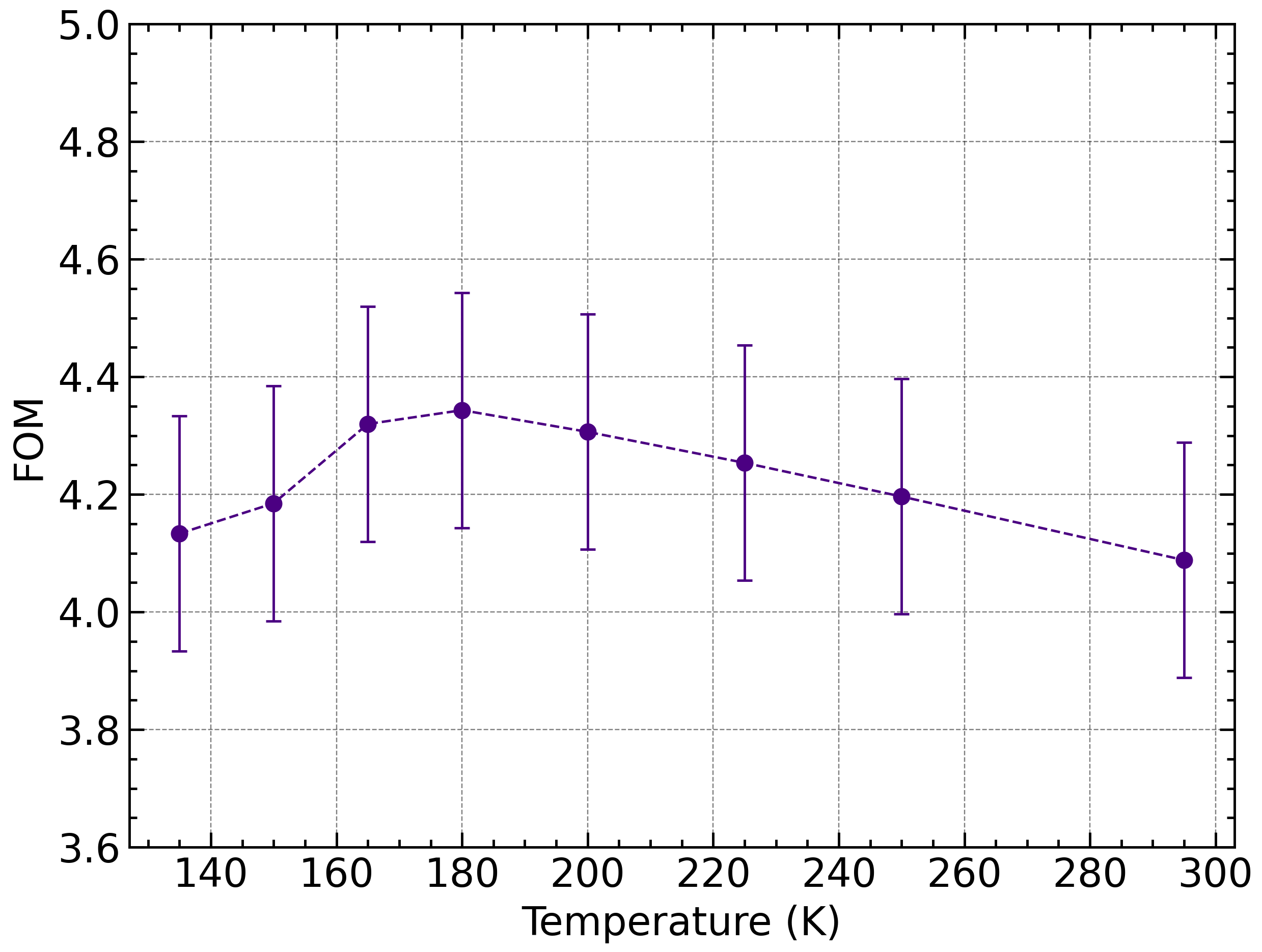}
    \caption{\label{FOMvsT} Temperature dependence of the Figure of Merit (\textit{FOM}) for pulse-shape discrimination measured with the Cs${_2}$ZrCl${_6}$ scintillating crystal in the temperature interval from RT down to 135~K.}
    
\end{figure}

In the framework of the PSD analysis, one can calculate the Figure Of Merit (\textit{FOM}), which is related to the goodness of the two peaks separation. It can be determined as:
\begin{eqnarray}
FOM = \frac{|\mu_{\alpha} - \mu_{\beta}|}{\sqrt{\sigma^2_{\alpha}+\sigma^2_{\beta}}}, 
\end{eqnarray}
where $\mu$ is the Gaussian mean and $\sigma$ is the standard deviation of the peaks of $\alpha$ and $\beta(\gamma)$ induced events in the mean-time parameter. It has been computed in the present case to be \textit{FOM} = 4.1 $\pm$ 0.2, value averaged in the energy interval (0.5--2.0)~MeV, in the $\gamma$-calibrated energy scale at RT. Due to the change in pulse shape with temperature, the experimentally determined \textit{FOM} varied with temperature as illustrated in figure~\ref{FOMvsT}. Between RT and 180~K, the \textit{FOM} is slightly raising, fluctuating between 4.1 $\pm$ 0.2 and 4.3 $\pm$ 0.2. Below temperatures of 180~K there was a reduction in \textit{FOM} down to 4.1 $\pm$ 0.2 at 135~K. It should be noted that despite this slight decrease in the \textit{FOM} parameter below 180~K coincides with a temperature range where the normalized relative intensities and decay time-constants for $\gamma$ events change significantly, it should be further confirmed in measurements with increased statistics in a 662~keV full absorption peak, reduced systematic uncertainties, and an enhanced temperature range down to 5~K.

\section{Discussion and conclusion}

Inorganic crystal scintillators can generally be considered as semiconductors with a wide band gap, and as it has been reported by Alig and Bloom~\cite{alig}, the energy required to generate an electron-hole pair by irradiation obey eq.~(\ref{ehproduction}):

\begin{equation}
    E = 2.73\cdot E_g + 0.55~\rm{eV}
    \label{ehproduction}
\end{equation}

where $E_g$ is the band gap of the semiconductor and the constant term accounts for energy lost to the crystalline lattice. These electron-hole pairs then thermalize, recombine and generate photons. The band gap of CZC has previously been evaluated as 2.78~eV by~\cite{BandGapCZC}. Later this value was reconsidered to be 4.67~eV \cite{zhang}. Therefore, the energy required to generate a single electron-hole pair in the CZC crystal is on average 13.3~eV, as determined by eq.~\ref{ehproduction}. This leads to the absolute light yield estimation at the level of 75,200~photons/MeV at RT, assuming ideal conditions where the photoluminescence quantum yield of luminescence centers and transport efficiency of the charge carriers both are equal to 100\% (or in other words, that each electron-hole pair produces one photon). Experimentally, the light yield of 50,300 $\pm$ 4,800~photons/MeV was evaluated in the table-top measurements with the CZC sample and exposed to 662~keV $\gamma$ quanta of the $^{137}$Cs source, exhibiting 67\% of the estimated absolute light yield value. At the same time, in measurements with an optical cryostat the light yield was evaluated as 53,300 $\pm$ 4,700~photons/MeV at RT under the same irradiation, exhibiting 71\% of its estimated absolute value. Both LY values evaluated from measurements in two different setups agreed well within their uncertainties.

The intensity of the scintillating light emission under $\alpha$ particles irradiation is rising when the CZC crystal is cooling down from RT to about 135~K reaching 19,700 $\pm$ 1,700~photons/MeV, and then the amount of the emitted light decreases to about 10,600 $\pm$ 900~photons/MeV at 75~K. While in the temperature interval from 75~K down to 5~K, it remains constant within uncertainties. It should be emphasized, that the present measurements confirms unusual behavior of the scintillating light intensity observed with the CZC crystal in~\cite{growth}, a decrease in the luminescence intensity as the temperature decreases from 125~K to 80~K, referred there as a “negative quenching”.

The light yield behavior under $\gamma$ quanta irradiation was studied only within RT--135~K temperature interval. It slightly rises from LY~=~53,300 $\pm$ 4,700~photons/MeV observed at RT, to LY~=~56,900 $\pm$ 5,000~photons/MeV measured at 165~K. Below 165~K the LY decreases to 53,800 $\pm$ 4,900~photons/MeV. The effect of reduction the scintillation light emission at different temperatures under different excitation type, 135~K under $\alpha$ particles irradiation and 165~K under $\gamma$ quanta irradiation, is needed to be further studied, as well as the extension of the temperature interval for the light yield evaluation for $\gamma$ quanta down to 5~K is strongly required.

It is well known that the luminescence properties of scintillation materials (in particular, emission maximum, spectrum shape and its bandwidth) can depend on temperature, see for instance~\cite{buryi2021correlation,zhang,bulyk2025luminescence}. Such changes can significantly affect the material scintillation efficiency at low temperatures. Therefore, we estimated the possible effect of emission spectrum variation with temperature on the experimentally detected light yield, considering previously published data on the temperature dependence of radio- and photo-emission spectra in CZC crystals~\cite{buryi2021correlation,zhang,bulyk2025luminescence}. The actual light yield of the CZC crystal at temperatures below 150~K could be about 10\% more than quoted here, in case one would account for the red-shift of the emission maximum (of about 15~nm) and narrowing the emission spectrum (FWHM reduction by about 25\%) at temperatures below 150~K with respect to the CZC luminescence characteristics at RT used in the present paper for QE and LY evaluations. Nevertheless, to make a precise correction, one would need to perform dedicated measurements of the luminescence emission spectra in a wide temperature range under irradiation by $\alpha$ particles and $\gamma$ quanta, to account for different ionization density of those two irradiation types that can lead to a significant variation in the emission maximum and its bandwidth. This could be a good topic for further studies with this scintillating material.

The quenching factor temperature trend, being a derivative parameter of the light yield temperature dependence of both $\alpha$ particles and $\gamma$ quanta, exhibits a smooth increase from QF~=~0.30 $\pm$ 0.04 at RT to QF~=~0.36 $\pm$ 0.05 at 135~K. Thus, for measurements with external $\alpha$ particles, or if processes involving the emission of internal $\alpha$ particles are being studied, a lower temperature is preferable.

The non-trivial trend of the scintillation light emission can be explained as follows. The conduction and valence bands in Cs${_2}$ZrCl${_6}$ are both formed by Zr 4d and Cl 3p electron orbitals, which leads to the concentration of the charge density in [ZrCl${_6}$]${^{2-}}$ octahedral clusters. At the same time, the nature of Cs${_2}$ZrCl${_6}$ lattice allows for structural distortions when the [ZrCl${_6}$]${^{2-}}$ cluster is in excited states, caused by stretching and shrinking of Zr–Cl bonds and accompanied by the change of bonds angle. These allow for a strong exciton localization and relaxation resulting in a high luminescence efficiency through STEs emission~\cite{zhang}. Moreover, the excited [ZrCl${_6}$]${^{2-}}$ clusters can populate the long-lived excited triplet states or a slightly more energetic the short-lived singlet state. At low temperatures, below 75~K, [ZrCl${_6}$]${^{2-}}$ clusters are predominantly filling the triplet excited states. The increase of the luminescence intensity in the temperature interval of 75--135~K could be associated with population of the singlet excited state from the triplet states through the reverse inter-system crossing process (RISC)~\cite{zhang}. The short-live singlet excited state is more energetic and causes an overall reduction of the decay time of scintillation emission with further temperature increase above 135~K.

Turning attention to the shape of scintillating pulses, one can derive following conclusions. While the average pulse for events induced by $\alpha$ particles is well-described by three decay time-constants (0.3, 2.5 and 11.8~$\mu$s at RT), the average pulse of events induced by $\gamma$ quanta is characterized by only two decay time-constants (1.3 and 11.5~$\mu$s at RT). Moreover, scintillating pulses induced by $\gamma$s exhibit a larger relative contribution of the slow decay time-component (84\% at RT) with respect to those induced by $\alpha$ particles (39\%), and this feature became more pronounced with the CZC crystal temperature decrease.

The quality of the pulse-shape discrimination, that could be quantified by the \textit{FOM} parameter, is slightly improving from \textit{FOM}~=~4.1 $\pm$ 0.2 at RT to \textit{FOM}~=~4.3 $\pm$ 0.2 at 180~K. This indicates an increase in signal shape variation induced by $\alpha$ particles and $\gamma$ quanta, caused by redistribution of the excitation energy between different luminescence centers responsible for scintillation processes with different decay time-constants and intensities. With further decrease of the sample temperature down to 135~K, the pulse-shape discrimination became less effective, as suggested by the \textit{FOM} parameter reduction. However, it should again be emphasized that future studies with CZC crystals should aim to reduce systematic uncertainties and to enhance the temperature range of pulse-shape studies for $\gamma$ induced events down to 5~K, to confidently exhibit the pronounced temperature dependence of the \textit{FOM} parameter.

Considering the light yield, pulse-shape discrimination capability and time-profiles of scintillation pulses, the use of the undopped CZC crystal at temperatures below 135~K is not practical, since the crystal will reach maximum of its scintillation performance at temperatures 135--165~K, followed by its acute deterioration below this temperature. Long-lasting scintillating pulses put also a significant limitation on the acceptable counting rate, especially at low temperatures. To mitigate this limitation, one could consider to dope CZC crystals with a highly efficient and fast activator such as Ce$^{3+}$ ions~\cite{kramer,jia}.

Further investigations of the Cs${_2}$ZrCl${_6}$ scintillating properties in a wide temperature range and under excitation by various types of irradiation is required. The light yield and scintillation pulse time-constants measured for $\gamma$ quanta below 135~K would complement values determined for $\alpha$ particles induced scintillation. Increasing the light collection efficiency through measurements of the crystalline sample with an integrating sphere at RT would provide a further clarification of the absolute light yield value of the bulk CZC crystal. This information could be used as an experimental correction factor to the light collection efficiency coefficients estimated in this study and further investigations.

\acknowledgments

The authors wish to recognize the support and funding provided by Queen's University and the Arthur B. McDonald Canadian Astroparticle Physics Research Institute. We would like to thank University of Toronto's Department of Physics for allocating time for the data analyst to preform various tasks. For the contribution of L.P. to this work: “This project has received funding from
the European Union’s Horizon 2020 research and innovation program under the Marie
Skłodowska-Curie Grant Agreement No.101029688 (ACCESS — Array of Cryogenic
Calorimeters to Evaluate Spectral Shape)”. S.N. expresses deep gratitude to Ioan Nahornyi for his continued support, critical discussions and interest in these research.

\clearpage

\bibliographystyle{unsrt}
\bibliography{Bibliography.bib}

\newpage

\end{document}